\documentclass[aps,prd,twocolumn,superscriptaddress,showpacs]{revtex4}

\usepackage{graphicx}
\usepackage{textcomp}
\usepackage{amssymb}
\usepackage{dcolumn}

\begin{document}


\title{Constraints on inelastic dark matter from XENON10}

\author{J.~Angle} \affiliation{Department of Physics, University of Florida, Gainesville, FL 32611, USA} 
\affiliation{Physics Institute, University of Z\"urich, Winterthurerstrasse 190, CH-8057,  Z\"urich, Switzerland}
\author{E.~Aprile} \affiliation{Department of Physics, Columbia University, New York, NY 10027, USA}
\author{F.~Arneodo} \affiliation{Gran Sasso National Laboratory, Assergi, L'Aquila, 67010, Italy}
\author{L.~Baudis} 
\affiliation{Physics Institute, University of Z\"urich, Winterthurerstrasse 190, CH-8057,  Z\"urich, Switzerland}
\author{A.~Bernstein} \affiliation{Lawrence Livermore National Laboratory, 7000 East Ave., Livermore, CA 94550, USA}
\author{A.~Bolozdynya} \affiliation{Department of Physics, Case Western Reserve University, Cleveland, OH 44106, USA}
\author{L.C.C.~Coelho} \affiliation{Department of Physics, University of Coimbra, R. Larga, 3004-516, Coimbra, Portugal}
\author{C.E.~Dahl} \affiliation{Department of Physics, Princeton University, Princeton, NJ 08540, USA}
\author{L.~DeViveiros} \affiliation{Department of Physics, Brown University, Providence, RI 02912, USA}
\author{A.D.~Ferella} \affiliation{Physics Institute, University of Z\"urich, Winterthurerstrasse 190, CH-8057,  Z\"urich, Switzerland} 
\affiliation{Gran Sasso National Laboratory, Assergi, L'Aquila, 67010, Italy}
\author{L.M.P.~Fernandes} \affiliation{Department of Physics, University of Coimbra, R. Larga, 3004-516, Coimbra, Portugal}
\author{S.~Fiorucci} \affiliation{Department of Physics, Brown University, Providence, RI 02912, USA}
\author{R.J.~Gaitskell} \affiliation{Department of Physics, Brown University, Providence, RI 02912, USA}
\author{K.L.~Giboni} \affiliation{Department of Physics, Columbia University, New York, NY 10027, USA}
\author{R.~Gomez} \affiliation{Department of Physics and Astronomy, Rice University, Houston, TX 77251, USA}
\author{R.~Hasty} \affiliation{Department of Physics, Yale University, New Haven, CT 06511, USA}
\author{L.~Kastens} \affiliation{Department of Physics, Yale University, New Haven, CT 06511, USA}
\author{J.~Kwong} \affiliation{Department of Physics, Princeton University, Princeton, NJ 08540, USA}
\author{J.A.M.~Lopes} \affiliation{Department of Physics, University of Coimbra, R. Larga, 3004-516, Coimbra, Portugal}
\author{N.~Madden} \affiliation{Lawrence Livermore National Laboratory, 7000 East Ave., Livermore, CA 94550, USA}
\author{A.~Manalaysay} \affiliation{Department of Physics, University of Florida, Gainesville, FL 32611, USA}  
\affiliation{Physics Institute, University of Z\"urich, Winterthurerstrasse 190, CH-8057,  Z\"urich, Switzerland}
\author{A.~Manzur} 
\affiliation{Department of Physics, Yale University, New Haven, CT 06511, USA}
\author{D.N.~McKinsey} \affiliation{Department of Physics, Yale University, New Haven, CT 06511, USA}
\author{M.E.~Monzani} \affiliation{Department of Physics, Columbia University, New York, NY 10027, USA}
\author{K.~Ni} \affiliation{Department of Physics, Yale University, New Haven, CT 06511, USA}
\author{U.~Oberlack} \affiliation{Department of Physics and Astronomy, Rice University, Houston, TX 77251, USA}
\author{J.~Orboeck} \affiliation{Department of Physics, RWTH Aachen University, Aachen, 52074, Germany}
\author{G.~Plante} \affiliation{Department of Physics, Columbia University, New York, NY 10027, USA}
\author{R.~Santorelli} \affiliation{Department of Physics, Columbia University, New York, NY 10027, USA}
\author{J.M.F.~dos~Santos} \affiliation{Department of Physics, University of Coimbra, R. Larga, 3004-516, Coimbra, Portugal}
\author{P.~Shagin} \affiliation{Department of Physics and Astronomy, Rice University, Houston, TX 77251, USA}
\author{T.~Shutt} \affiliation{Department of Physics, Case Western Reserve University, Cleveland, OH 44106, USA}
\author{P.~Sorensen}
\email{pfs@llnl.gov}
\affiliation{Lawrence Livermore National Laboratory, 7000 East Ave., Livermore, CA 94550, USA}
\author{S.~Schulte} \affiliation{Department of Physics, RWTH Aachen University, Aachen, 52074, Germany}
\author{C.~Winant} \affiliation{Lawrence Livermore National Laboratory, 7000 East Ave., Livermore, CA 94550, USA}
\author{M.~Yamashita} 
\affiliation{Department of Physics, Columbia University, New York, NY 10027, USA}

\collaboration{XENON10 Collaboration}\noaffiliation
\date{\today}

\begin{abstract}
It has been suggested that dark matter particles which scatter inelastically from detector target nuclei could explain the apparent incompatibility of the DAMA modulation signal (interpreted as evidence for particle dark matter) with the null results from CDMS-II and XENON10.  Among the predictions of inelastically interacting dark matter are a suppression of low-energy events, and a population of nuclear recoil events at higher nuclear recoil equivalent energies.  This is in stark contrast to the well-known expectation of a falling exponential spectrum for the case of elastic interactions.  We present a new analysis of XENON10 dark matter search data extending to E$_{nr}=75$~keV nuclear recoil equivalent energy.  Our results exclude a significant region of previously allowed parameter space in the model of inelastically interacting dark matter.  In particular, it is found that dark matter particle masses $m_{\chi}\gtrsim150$~GeV are disfavored.
\end{abstract}

\maketitle

\section{Introduction}
Astrophysical evidence indicates that 23\% of the mass of the universe is in the form of non-baryonic dark matter \cite{2008hinshaw,2007jee,2006clowe}.  A well-motivated dark matter candidate particle is found in supersymmetric extensions to the Standard Model, in which the lightest supersymmetric particle (LSP) is stable.  A cosmologically interesting relic density of Weakly Interacting Massive Particles (WIMPs) arises from rather general arguments \cite{1996jungman}, which implies that the LSP could be dark matter.  The open question of the expected mass and cross-section of particle dark matter is being addressed by numerous direct and indirect detection experiments \cite{gaitskell2004,baudis2006}.  The nature of the dark matter particle, and its coupling to standard model particles has been the subject of much recent theoretical investigation \cite{2009arkanihamed, 2009cui,2009alves}.

\subsection{The XENON10 detector}
The XENON10 detector is a liquid xenon time-projection chamber with an active target mass of 13.7~kg.  It was designed to directly detect galactic dark matter particles which scatter off xenon nuclei.  Typical velocities of halo-bound dark matter particles are of order 10$^{-3}$c.  This leads to a a featureless exponential recoil energy spectra for spin-independent elastic scattering of dark matter particles on a xenon target.  In the case of a 100~GeV dark matter particle mass, the predicted elastic scattering recoil energy spectrum falls by an order of magnitude from $0-30$~keV nuclear recoil equivalent energy (keVr) \cite{1996lewin,1996jungman}.  

A particle interaction in liquid xenon creates both excited and ionized xenon atoms \cite{1978kubota}, which react with the surrounding xenon atoms to form excimers.  The excimers relax on a scale of $10^{-8}$~s with the release of scintillation photons.  This prompt scintillation light is detected by 88 photo-multiplier tubes and is referred to as the $S1$ signal.  An external electric field ($E_d=0.73$~kV/cm) across the liquid xenon target causes a large fraction of ionized electrons to be drifted away from an interaction site.  The electrons are extracted from the liquid xenon and accelerated through a few mm of xenon gas by a stronger electric field  ($\sim10$~kV/cm), creating a secondary scintillation signal.  This scintillation light is detected by the same photo-multiplier tubes, is proportional to the number of ionized electrons and is referred to as $S2$.  

The XENON10 detector discriminates between electron recoil background and the expected nuclear recoil signal from the scattering of a dark matter particle, via the distinct ratio of proportional ($S2$) to primary ($S1$) scintillation for each type of interaction.  The XENON10 collaboration previously reported WIMP-nucleon exclusion limits for spin-independent \cite{2008xenon10SI} and spin-dependent \cite{2008xenon10SD} elastic scattering.


\subsection{Inelastic dark matter}
In stark contrast to elastic scattering \cite{1996lewin,1996jungman}, a model of inelastic interactions of dark matter (iDM) predicts \cite{2001sw,2005sw} a suppression of low-energy nuclear recoils.  The recoil energy spectrum peaks near $40$~keVr, for a 100~GeV dark matter particle incident on a xenon target (and standard halo assumptions).  As explained in \cite{2005sw}, this feature is a result of the mass difference $\delta$ between the proposed ground and excited states of the dark matter particle.  The value of $\delta$ is unknown and is a free parameter in the model, subject to the physical constraint that $\delta \gtrsim 170$~keV would kinematically forbid any scattering from dark matter particles, given an expected particle mass $m_{\chi}=100$~GeV and a galactic escape velocity $v_{esc}=500$~km~s$^{-1}$.  In the limit $\delta\rightarrow0$, the iDM model reduces to the usual elastic scattering case.

At present, the iDM model is comfortably consistent \cite{2009chang} with all reported null results (including the recent CDMS results \cite{2009cdms}) as well as the claimed detection from DAMA \cite{2008dama}.  Interpretation of the XENON10 results in \cite{2009chang} was limited by the fact that in \cite{2008xenon10SI}, data were not published beyond 45~keVr.  Since the focus in \cite{2008xenon10SI} was on elastic interactions, that analysis was optimized for events with recoil energies in the range $4.5-26.9$~keVr.  This paper presents a new analysis of the XENON10 dark matter search data, extending to the energy range of interest for inelastic dark matter scattering.  In doing so, we find new constraints on allowed parameter space in the iDM model.  

\section{Predicted Event Rates}
We calculate predicted differential event rates as a function of nuclear recoil energy in XENON10 and DAMA following \cite{2009chang}, as 
\begin{equation} \label{rate_eq}
\frac{dR}{dE_{nr}} = N_T M_N \frac{\rho_{\chi} \sigma_n}{2 m_{\chi} \mu_{ne}^2}
     A^2 
     F^2(E_{nr}) \int_{\beta_{min}}^{v_{esc}} \frac{f(v)}{v} dv.
\end{equation}
The number of target nuclei in the detector is $N_T$, the mass of the target nucleus is $M_N$ and its atomic number is $A$.  We assume the standard local dark matter density $\rho_{\chi} = 0.3$~GeV~cm$^{-3}$, with dark matter particle mass $m_{\chi}$ and cross section (per nucleon) $\sigma_n$.  Therefore, the reduced mass $\mu_{ne}$ is for the nucleon $-$ dark matter particle system.  

The nuclear form factor $F(E_{nr})$ accounts for a loss of coherence as momentum transfer to the nucleus increases.  We use the Helm form factor parameterization $F(E_{nr}) = 3j_1(qr_n)/qr_n \cdot \exp(-(qs)^2/2)$.  We take the effective nuclear radius $r_n = \sqrt{r_0^2-5s^2}$, with $r_0=1.2A^{1/3}$~fm and the skin thickness $s=1$~fm.  The momentum transfer to the nucleus is just $q=\sqrt{2M_NE_{nr}}$.  It was shown in \cite{2006duda} that $F^2$ using this parameterization differs by 8\%~(13\%) at 30~keVr (60~keVr) compared with other reasonable parameterizations.  As noted in \cite{2009chang}, this can lead to differences of up to 25\% in the predicted rate.  For consistency, we therefore use this same form factor to calculate parameter space consistent with DAMA and excluded by XENON10.  

We assume the dark matter velocity distribution to be Maxwellian, and perform the integration over $f(v)/v$ following \cite{2006savage}.  As in \cite{2006savage}, we take the rotational speed of the local standard of rest to be $v_0=220$~km~s$^{-1}$. We used a weighted average of the velocity distribution over the period October 2006 $\rightarrow$ February 2007, corresponding to when the data were acquired.  The lower limit of the integral in Eq. \ref{rate_eq} is $\beta_{min} = \sqrt{1/(2m_NE_{nr})}(m_NE_{nr}/\mu+\delta)$, which reduces to the usual $v_{min}$ (as in \cite{1996lewin}, for example) in the case $\delta\rightarrow0$. The upper limit is set by the galactic escape velocity, which we take to be $v_{esc}=500$~km~s$^{-1}$ as in \cite{2009chang}. Uncertainties in exclusion limits arising from assumptions about $v_{esc}$ are discussed further in Sec. \ref{iDMlimits}.

\section{Event acceptance} \label{eventAcc}
\begin{figure}[h]
\centering
\includegraphics[width=0.49\textwidth]{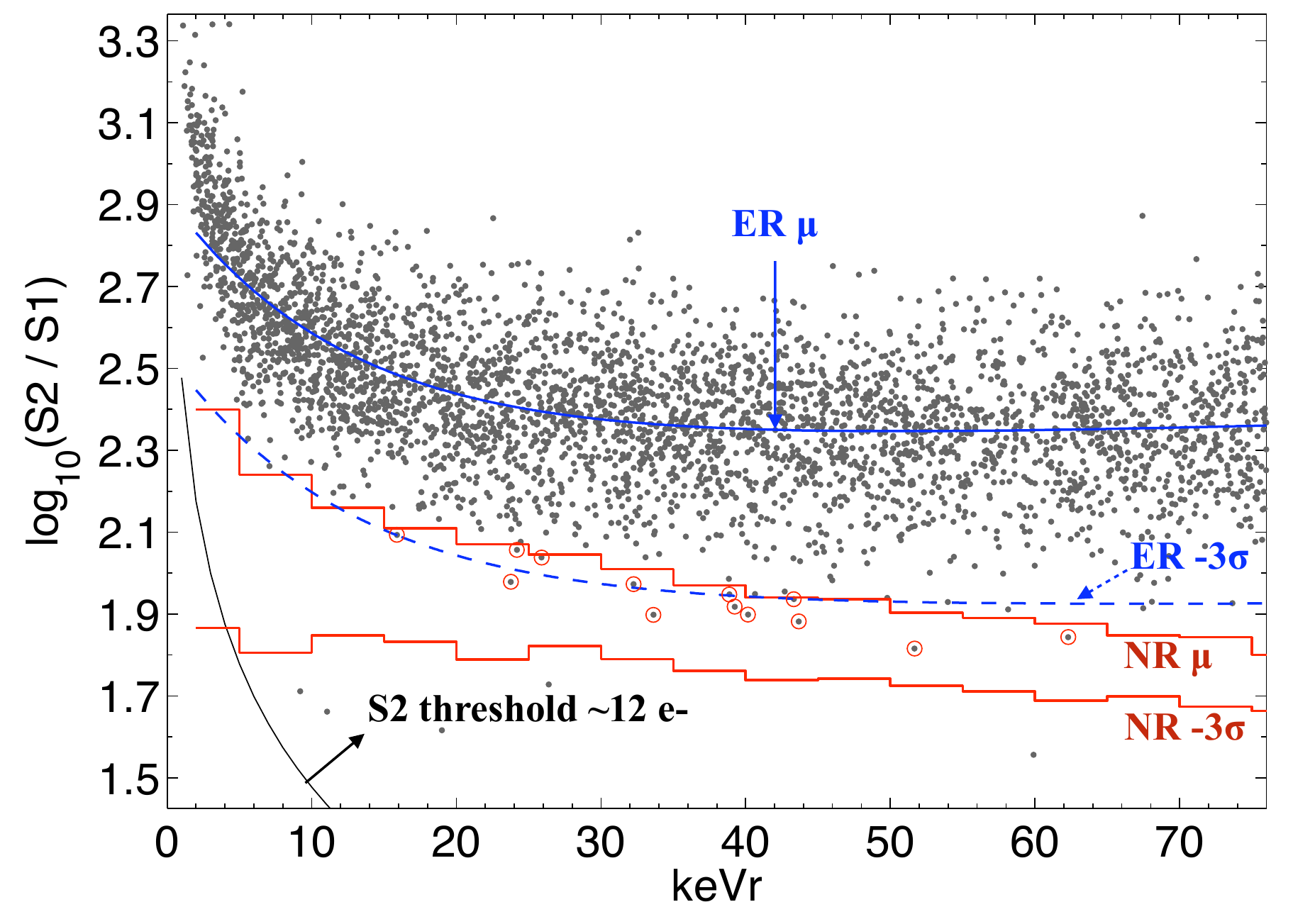}
\caption{(color online) Events remaining after re-analysis of 58.6 live days of dark matter search data.  The dark matter acceptance box is bounded by the stair-step lines indicating the centroid and $-3\sigma$ bounds obtained from fits to the neutron calibration data.  The fitted electron recoil centroid is shown as a solid blue curve, and the $-3\sigma$ contour is shown dashed.  The 13 events remaining in the acceptance box are circled.  An $S2$ software threshold of 12 electrons was imposed in the analysis.}
\label{fig1}
\end{figure}

Figure \ref{fig1} shows the results of a re-analysis of 58.6 live days of dark matter search data already described in \cite{2008xenon10SI,2008xenon10SD}, with the fiducial target mass unchanged at 5.4~kg.  While the previous analyses were performed blind, the present work introduces two new software cuts (discussed in Sec. \ref{ppc} and Sec. \ref{fprompt}) and so cannot be considered as blinded.  The hardware trigger efficiency was verified to be greater than $99\%$ for 4 electrons in the $S2$ signal, and a software threshold of 12 electrons was imposed in the analysis.

The recoil energy range is extended to the region relevant for inelastic dark matter scattering, and is quoted in keVr.  The energy scale is calculated from the primary scintillation signal as $E_{nr}=S1/(L_y~\mathcal{L}_{eff}) \cdot (S_e/S_n)$, with $L_y=3.0\pm0.14$~photo-electrons/keVee the measured light yield for 122~keV photons.  The scintillation quenching of electron and nuclear recoils due to the electric field $E_d$ are $S_e=0.54\pm0.02$ \cite{2005aprile} and $S_n=0.95\pm0.05$ \cite{2009manzur}.  $\mathcal{L}_{eff}$ is the relative scintillation efficiency of liquid xenon for nuclear recoils.  Due to the varied results from experimental determinations of $\mathcal{L}_{eff}$ \cite{2005aprile,2009sorensen,2009aprile,2009manzur,2006chepel}, we show our results in Fig. \ref{fig1} assuming a constant $\mathcal{L}_{eff}=0.19$.  From this, we find $1.00\pm0.15~S1$~photo-electron/keVr.  Prior to calculating exclusion limits, appropriate values of keVr for each remaining event were obtained based on recent measurements of $\mathcal{L}_{eff}$ \cite{2009sorensen,2009aprile,2009manzur}.

Background rejection is obtained for each event from the ratio of proportional to primary scintillation, written as $\mbox{log}_{10}(S2/S1)$.  An acceptance box for nuclear recoils was defined from the nuclear recoil calibration data, following the procedure described in \cite{2008xenon10SI}.  This is indicated by the stair-step lines in Fig. \ref{fig1}, and the acceptance $A_{nr}$ is listed in Table \ref{table1}.  Additional background rejection is accomplished with software cuts on $S1$ photo-multiplier tube optical distributions (described in Sec. \ref{ppc}), and on the $S1$ decay time distribution (described in Sec. \ref{fprompt}).  

The 13 events from the dark matter search data that remained in the acceptance box after all cuts are indicated in Fig. \ref{fig1} by red circles.  No additional events remained in or below the acceptance box in the energy range $75-250$~keVr, and in the interest of clarity this is not shown in Fig. \ref{fig1}.  The acceptance remains $A_{nr}\sim 0.50$ in this higher energy range.  The 5 events below the acceptance box in Fig. \ref{fig1} are consistent with the false single scatter pathology described in Sec. \ref{backgrounds}.

\begin{table}[h]
\centering
\caption{Software cut acceptance for nuclear recoils $\epsilon_p$ and $\epsilon_f$ (discussed in Sec. \ref{ppc} and Sec. \ref{fprompt}), and the nuclear recoil band acceptance A$_{nr}$, as a function of nuclear recoil energy.  The expected number of events N$_{stat}$ in the acceptance box is determined from the number of detected events N$_{evts}$ and the Gaussian width of the electron recoil $\mbox{log}_{10}(S2/S1)$ distribution.}
\begin{tabular}{cccccc}
\\
\toprule
E$_{nr}$ (keVr) & ~$\epsilon_p$~ & ~$\epsilon_{f}$~ & ~A$_{nr}$~ & ~N$_{evts}$~ & ~N$_{stat}~$  \\
\hline
02-05 & 1.00 & 0.91 & 0.47 & 228 & 0.3$^{+0.2}_{-0.1}$ \\ 
05-10 & 0.92 & 0.90 & 0.46 & 408 & 0.3$^{+0.2}_{-0.1}$ \\ 
10-15 & 0.83 & 0.91 & 0.46 & 351 & 0.9$^{+0.4}_{-0.3}$ \\ 
15-20 & 0.67 & 0.89 & 0.48 & 269 & 1.1$^{+0.4}_{-0.3}$ \\ 
20-25 & 0.62 & 0.91 & 0.48 & 265 & 1.1$^{+0.4}_{-0.4}$ \\ 
25-30 & 0.59 & 0.92 & 0.46 & 259 & 0.9$^{+0.4}_{-0.3}$ \\ 
30-35 & 0.61 & 0.90 & 0.48 & 267 & 1.4$^{+0.5}_{-0.4}$ \\ 
35-40 & 0.64 & 0.92 & 0.48 & 252 & 0.7$^{+0.3}_{-0.3}$ \\ 
40-45 & 0.65 & 0.92 & 0.49 & 239 & 0.3$^{+0.1}_{-0.1}$ \\ 
45-50 & 0.56 & 0.92 & 0.52 & 218 & 0.2$^{+0.1}_{-0.1}$ \\ 
50-55 & 0.64 & 0.92 & 0.54 & 216 & 0.0$^{+0.0}_{-0.0}$ \\ 
55-60 & 0.63 & 0.90 & 0.48 & 167 & 0.1$^{+0.1}_{-0.0}$ \\ 
60-65 & 0.65 & 0.89 & 0.51 & 202 & 0.1$^{+0.0}_{-0.0}$ \\ 
65-70 & 0.63 & 0.89 & 0.48 & 203 & 0.0$^{+0.0}_{-0.0}$ \\ 
70-75 & 0.67 & 0.94 & 0.53 & 198 & 0.0$^{+0.0}_{-0.0}$ \\ 
\botrule
\end{tabular}

\label{table1}
\end{table}

\subsection{Poisson likelihood} \label{ppc}
As explained in \cite{2008xenon10SI}, a residual population of background events were observed in the electron recoil calibration data.  These events tend to populate the nuclear recoil signal region.  The origin of these events is discussed in Sec. \ref{backgrounds}.  They were targeted with high efficiency by calculating the Poisson likelihood of the $S1$ hit pattern of each event.  For each event, the probability $\alpha_i$ that the $i^{th}$ photo-multiplier tube registered a photo-electron (resulting from the $S1$ primary scintillation signal) is a function of the event vertex, so $\alpha_i = \alpha_i(x,y,z)$.  As described in \cite{2008xenon10SI}, the event $(x,y)$ coordinates are determined by the pattern of proportional scintillation on the top photo-multiplier tube array, while the $z$ coordinate is measured from the time delay between the primary and proportional scintillation.

The distribution of $\alpha_i$ for each photo-multiplier tube was measured from calibration data, obtained after introducing neutron-activated xenon into the XENON10 detector.  As described in \cite{2007ni}, this produced an internal, homogeneous source of 164 keV gamma rays from the de-excitation of $^{131m}$Xe.  For each event, we then calculated the Poisson probability $p_i$ of obtaining the observed hit-pattern, given the expectation $\alpha_i$.  Two cut parameters were defined as $\mathcal{P}_{b,t}=\mbox{log}_{10}(\Sigma~\alpha_i / p_i)$.  In the case of $\mathcal{P}_b$ ($\mathcal{P}_t$), the sum ran over only the bottom 41 (top 47) photo-multiplier tubes.  Linear cut bands were optimized for $\mathcal{P}_{b}$ and $\mathcal{P}_{t}$ in order to remove leakage events in the electron recoil calibration data.  This is discussed further in Sec. \ref{backgrounds}.  The acceptance for nuclear recoils $\epsilon_p$ for the combined $\mathcal{P}_{b,t}$ software cuts is shown in Table \ref{table1}.  

\subsection{Primary scintillation pulse shape} \label{fprompt}
\begin{figure}[h]
\centering
\includegraphics[width=0.45\textwidth]{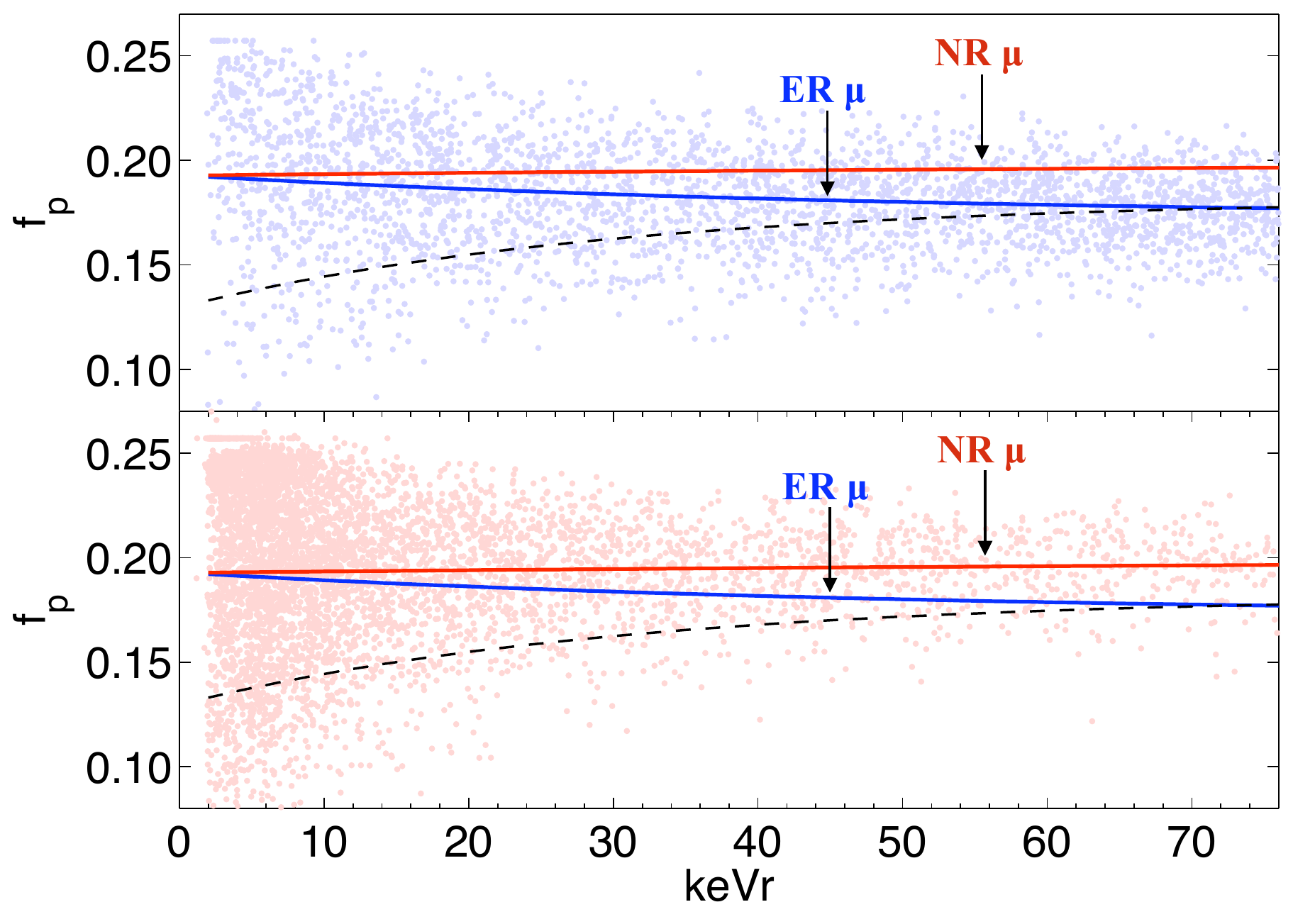}
\caption{(color online) {\bf (top)} The distribution of $f_{p}$ for electron recoil calibration data.  The black dashed line indicates the discrimination cut bound on this parameter. {\bf (bottom)} The distribution of $f_{p}$ for nuclear recoil calibration data.  The acceptance $\epsilon_f$ of the cut is listed in Table \ref{table1}.}
\label{fig2}
\end{figure}

It is well known that the primary scintillation light from nuclear recoils and electron recoils in liquid xenon exhibit distinct decay times corresponding to the preferentially excited singlet and triplet states of the Xe$_2^*$ dimer \cite{1983hitachi}.  The resulting pulse shape discrimination is significantly less powerful than discrimination based on $\mbox{log}_{10}(S2/S1)$ (see \cite{2005alner}, for example).  This is due to the modest separation between the lifetime of the triplet (27~ns) and singlet ($\lesssim4$~ns) states \cite{1999doke}.  At small recoil energies, Poisson fluctuations in the number and arrival times of photo-electrons increase the width of the distribution, as shown in Fig. \ref{fig2}.  It is also worth mentioning that the 105~MHz clock speed of the ADC used in XENON10 is not optimal for characterizing xenon scintillation pulse shapes.

Still, a modest benefit can be obtained by parameterizing the primary scintillation $S1=S1(t)$ in terms of its prompt fraction (following \cite{2009kwong}),
\begin{equation}
f_{p} = \frac{\int_{t_i}^{t_0+t_w} S1(t) dt}{\int_{t_i}^{t_f} S1(t) dt}.
\end{equation}
The point at which a primary scintillation pulse rises to 10\% of its maximum was defined as $t_0$, $t_i=t_0-5$, $t_f=t_0+30$ and the prompt window $t_w=4$.  Times are quoted in units of $1/105~\mu$s ADC samples, so $t_w\approx38$~ns.  As discussed in \cite{2009kwong}, the optimal value of $t_w$ depends on the electronics and the size of the detector.  The discrimination was only slightly decreased if we chose $t_w=3$, while $t_w<3$ or $t_w>4$ showed significantly weaker performance.  The prompt fraction $f_p$ obtained for neutron and electron recoil calibration data are shown in Fig. \ref{fig2}, along with curves indicating the centroid of each distribution.  A cut which maintains $\epsilon_f\simeq0.9$ acceptance for nuclear recoils is indicated by the black dashed curve.  The acceptance as a function of energy is shown in Table \ref{table1}.

The electron recoil rejection efficiency obtained from this cut rises monotonically from $R_{er}=0.15$ at $E_{nr}=10$~keVr to $R_{er}=0.5$ at $E_{nr}=70$~keVr.  This level of rejection is significantly weaker than the rejection already obtained from discrimination based on $\mbox{log}_{10}(S2/S1)$.  However, it is an independent channel for characterizing the electromagnetic background, and is especially useful at higher energies.  

\section{Electromagnetic Backgrounds} \label{backgrounds}
The signal acceptance box in Fig. \ref{fig1} is subject to contamination (``leakage'') from the electromagnetic background.  This background divides naturally into two components:  statistical leakage, and (following the nomenclature of \cite{2008xenon10SI}) anomalous leakage.  Predictions for the former were obtained from Gaussian fits to the $\mbox{log}_{10}(S2/S1)$ distributions observed in calibration data, and are shown in Table \ref{table1} (N$_{stat}$).  Predictions for the latter, which will be shown to arise from mis-identified multiple scatter events, are not tabulated.  This is due to the small number of such events in the calibration data sample.  Instead, we calculate the predicted number of events for the energy range $E_{nr}<75$~keVr.

Multiple scatter events in the active 13.7~kg region of the XENON10 detector are easy to identify, because each scatter vertex creates a separate proportional scintillation ($S2$) pulse.  However, there are also 8.7~kg of xenon entirely outside the active region, and $1.2$~kg of xenon in the region between the bottom photo-multiplier tube array and the cathode grid.  A rendering of the detector which shows these regions can be found in \cite{2009sorensen}.  The 1.2~kg of xenon below the cathode grid contribute $S1$ but not $S2$, due to the reversed direction of the electric field $E_d$ in this region.  A multiple scatter event with one vertex in this region and one vertex in the central 5.4~kg target would always $-$ in the absence of the $\mathcal{P}_{b,t}$ (or similar) characterization $-$ be tagged as a valid single scatter.  

The 8.7~kg of xenon in the outer region were expected to be entirely passive, i.e. neither $S1$ nor $S2$ would be observed from energy depositions in this region.  However, a ray-tracing Monte Carlo simulation \cite{2007gomez} showed that the light collection efficiency in this region ranged as high as 15\% (in a small region near the lower photo-multiplier tube array), with a mean of about 1\%.  

Both the reverse field region and the outer region allow the possibility that a multiple scatter background event is tagged as a single scatter event (a ``false single scatter'').  Some fraction of the $S1$ from the additional scatter may be collected, while none of the $S2$ will be collected.  This implies that many false single scatters will have anomalously low values of $\mbox{log}_{10}(S2/S1)$, compared with genuine single scatter events.  The lower bound in $\mbox{log}_{10}(S2/S1)$ for false single scatter events is set by the $S2$ software threshold, so these events may fall below the nuclear recoil acceptance box as well as in it. This is observed in electron recoil calibration data.

\begin{figure}[h]
\centering
\includegraphics[width=0.45\textwidth]{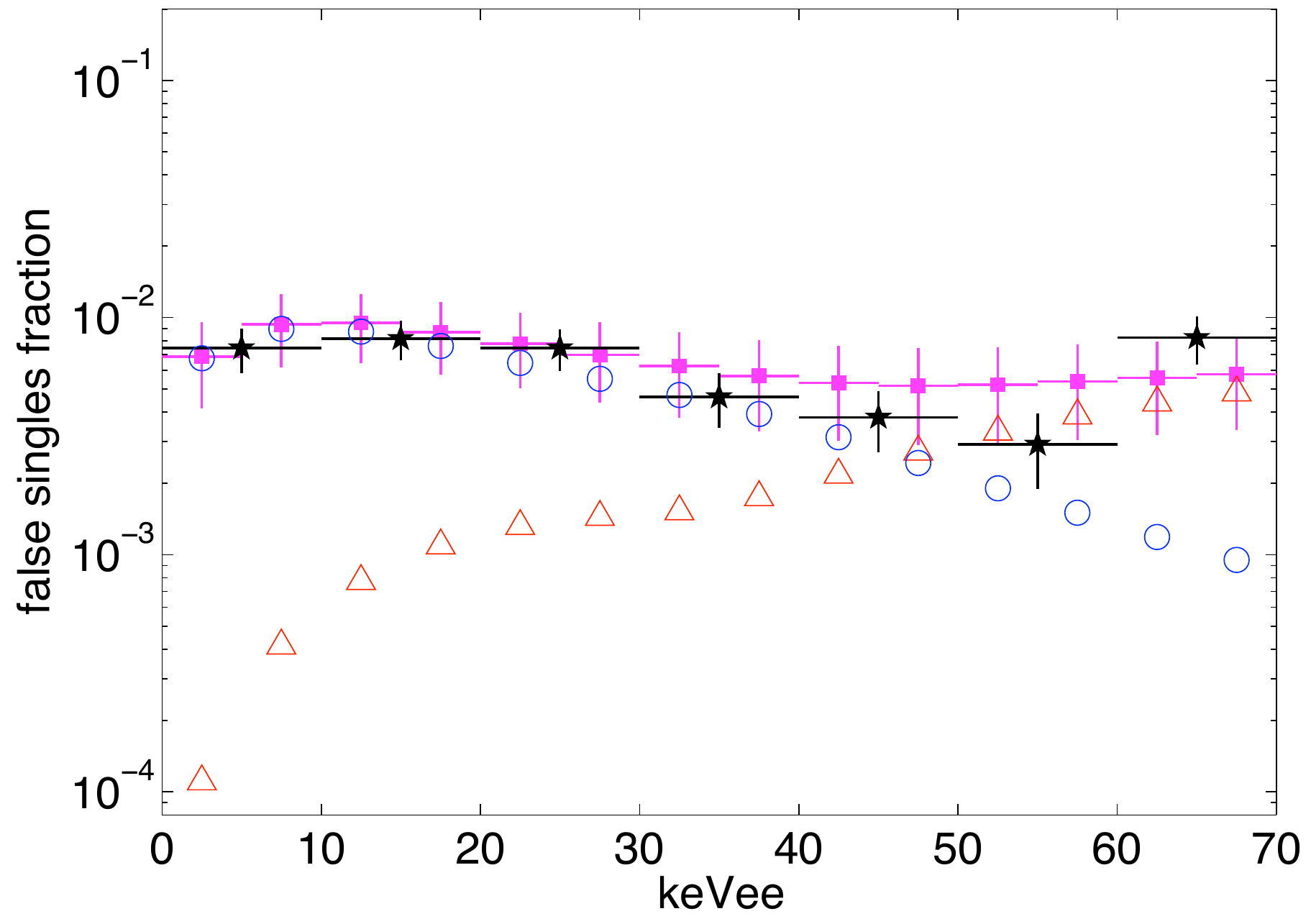}
\caption{(color online) The fraction of leakage events remaining among single gamma ray scatters in the electron recoil calibration data set (black stars).  Also shown are the Monte Carlo prediction for false single scatters, with an additional scatter below the cathode grid (red triangles), or in the outer 8.7~kg of xenon (blue circles).  The sum of the two Monte Carlo components is indicated by pink squares.  Note that the $x$ axis is electron (not nuclear) recoil equivalent energy;  20~keVee = 76~keVr.}
\label{fig3}
\end{figure}

Figure \ref{fig3} (black stars) shows the fraction of leakage events remaining among single scatters in an electron recoil calibration data set, obtained using a $^{137}$Cs source.  This fraction was calculated prior to applying cuts based on $\mathcal{P}_{b,t}$ or $f_p$, and only for events tagged as single scatters in the 5.4~kg target.  An event was considered to be a leakage event if its $\mbox{log}_{10}(S2/S1)$ value fell below the $-3\sigma$ contour for electron recoils, which is shown as a dashed line in Fig. \ref{fig1}.  To obtain the correct electron equivalent energy scale, the data were corrected for the measured scintillation yield as a function of energy \cite{2008sorensen}.

A Monte Carlo simulation of the XENON10 detector response to this electron recoil calibration was used to obtain a prediction for the false single scatter fraction.  The result is shown in Fig. \ref{fig3}, divided into two components:  false singles with an additional scatter below the cathode grid (red triangles), and false singles with an additional scatter in the outer 8.7~kg of xenon (blue circles).  The sum of the two components is also shown (pink squares).  The simulation was corrected to account for the predicted light collection efficiency in each region \cite{2007gomez}; for the scintillation quenching \cite{2005aprile}, based on the electric field strength in each region \cite{2007gomez}; and for the observed energy resolution of the detector.  

For each simulated false single scatter event, a value of $\mbox{log}_{10}(S2/S1)$ was calculated.  This was done by assuming that each individual scatter vertex had a typical electron recoil $\mbox{log}_{10}(S2/S1)$ value, given by the centroid of the distribution (Fig. \ref{fig1}, solid blue line).  The simulation prediction in Fig. \ref{fig3} only includes false single scatters whose $\mbox{log}_{10}(S2/S1)$ was low enough that the event would be characterized as leakage.  This criteria was met by 26\% of events with an additional scatter below the cathode grid, and by 5\% of events with an additional scatter in the outer 8.7~kg of passive xenon.

It is clear from Fig. \ref{fig3} that the observed leakage fraction and the predicted false single scatter fraction are very similar in the calibration data.  Making this comparison for the dark matter search data is significantly more involved (and is a work in progress \cite{2009deviveiros}), since it requires detailed modeling of the location and isotopes which contribute to the electromagnetic background.  

From the electron recoil calibration data, the rejection efficiency for false single scatter events in the nuclear recoil acceptance box was calculated after all cuts (as described in Sec. \ref{ppc} and Sec. \ref{fprompt}) to be R$_{leak}=0.851\pm0.056$.  If this same level of rejection applies to background events in the dark matter search data, there should remain $12.0\pm4.5$ false single scatter events in the acceptance box in Fig. \ref{fig1}.  

If we account for statistical leakage (N$_{stat}$ from Table \ref{table1}), the 13 remaining events in Fig. \ref{fig1} would be reduced to 7.5 events.  The rejection efficiency for the electron recoil calibration data in this case is R$_{leak}=0.908\pm0.052$.  If this same level of rejection applies to background events in the dark matter search data, we predict $6.5\pm4.0$ events remaining in the acceptance box in Fig. \ref{fig1}.  In either case, the number of remaining potential signal events are consistent with expected background.

\section{dark matter exclusion limits} \label{iDMlimits}
\begin{figure*}[p]
\includegraphics[width=0.42\textwidth]{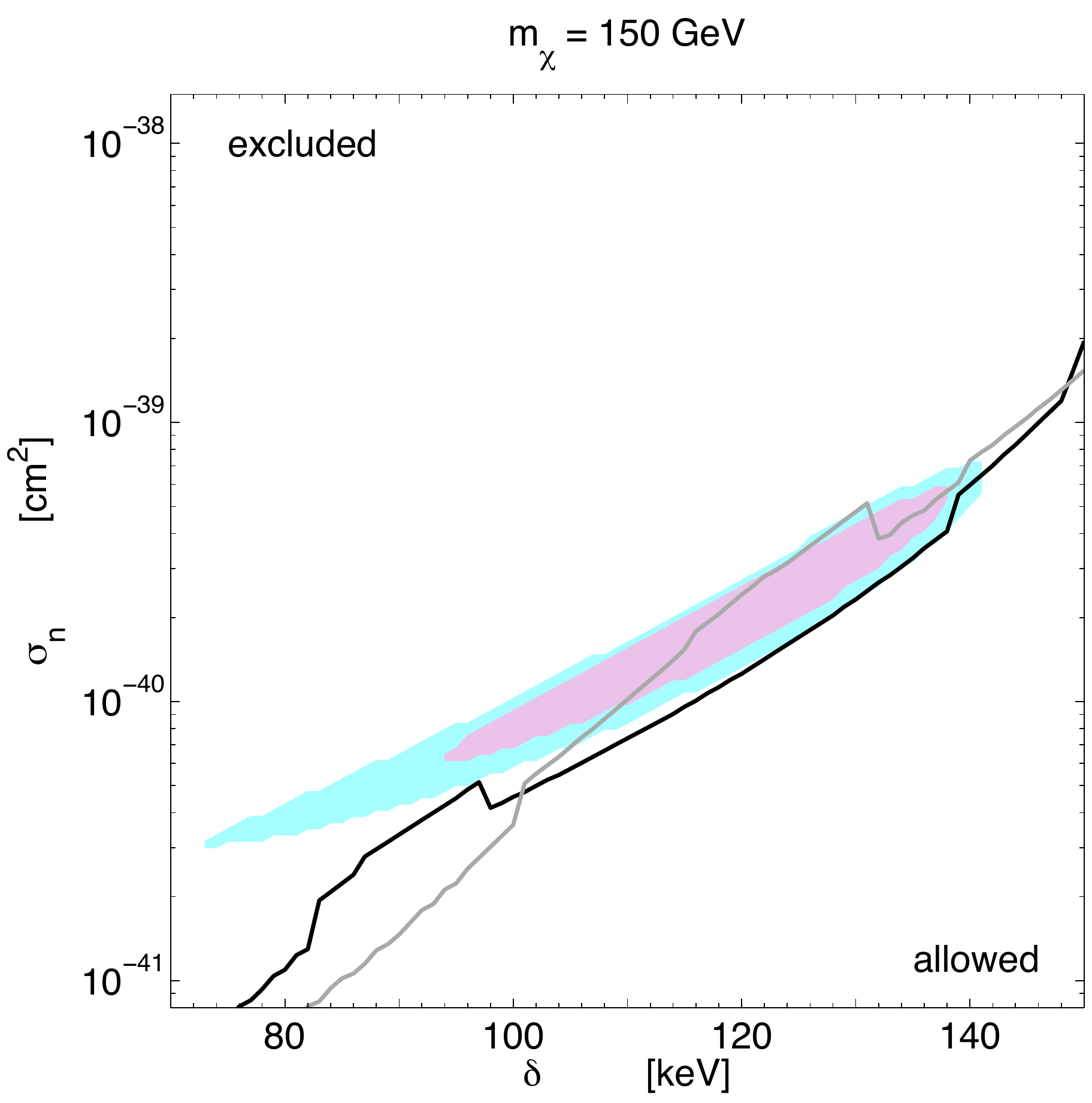}
\includegraphics[width=0.42\textwidth]{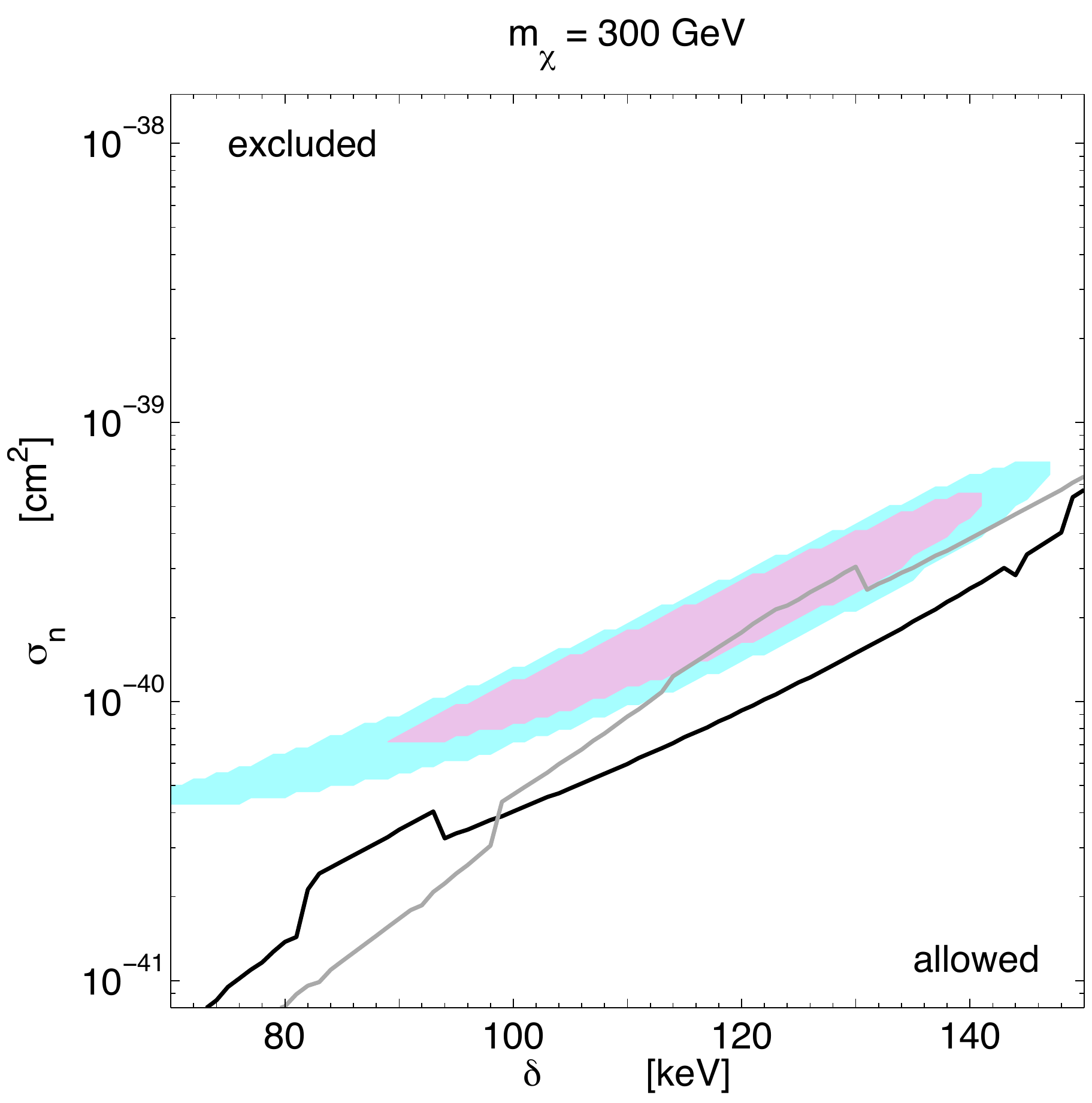}
\includegraphics[width=0.42\textwidth]{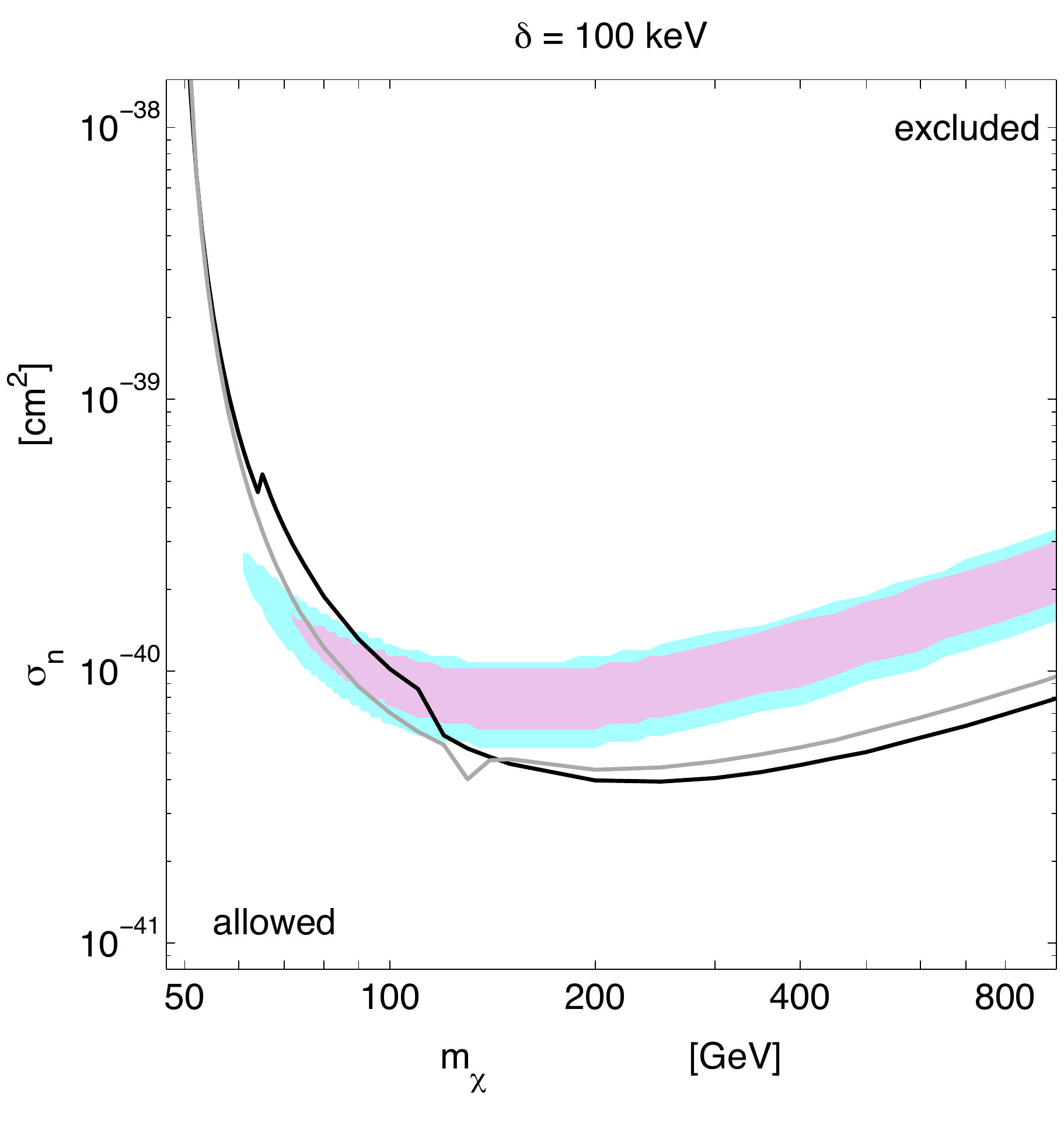}
\includegraphics[width=0.42\textwidth]{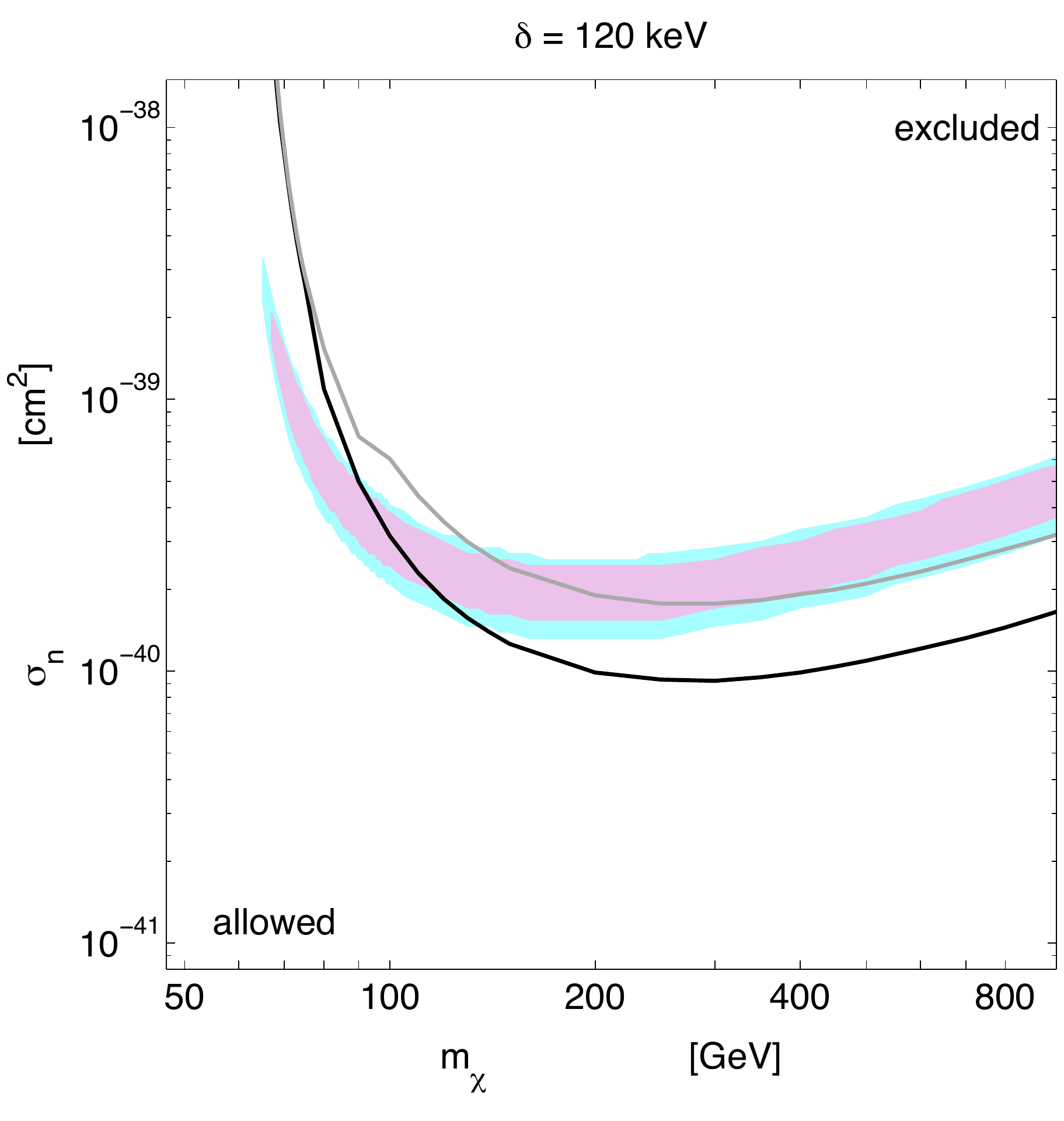}
\includegraphics[width=0.42\textwidth]{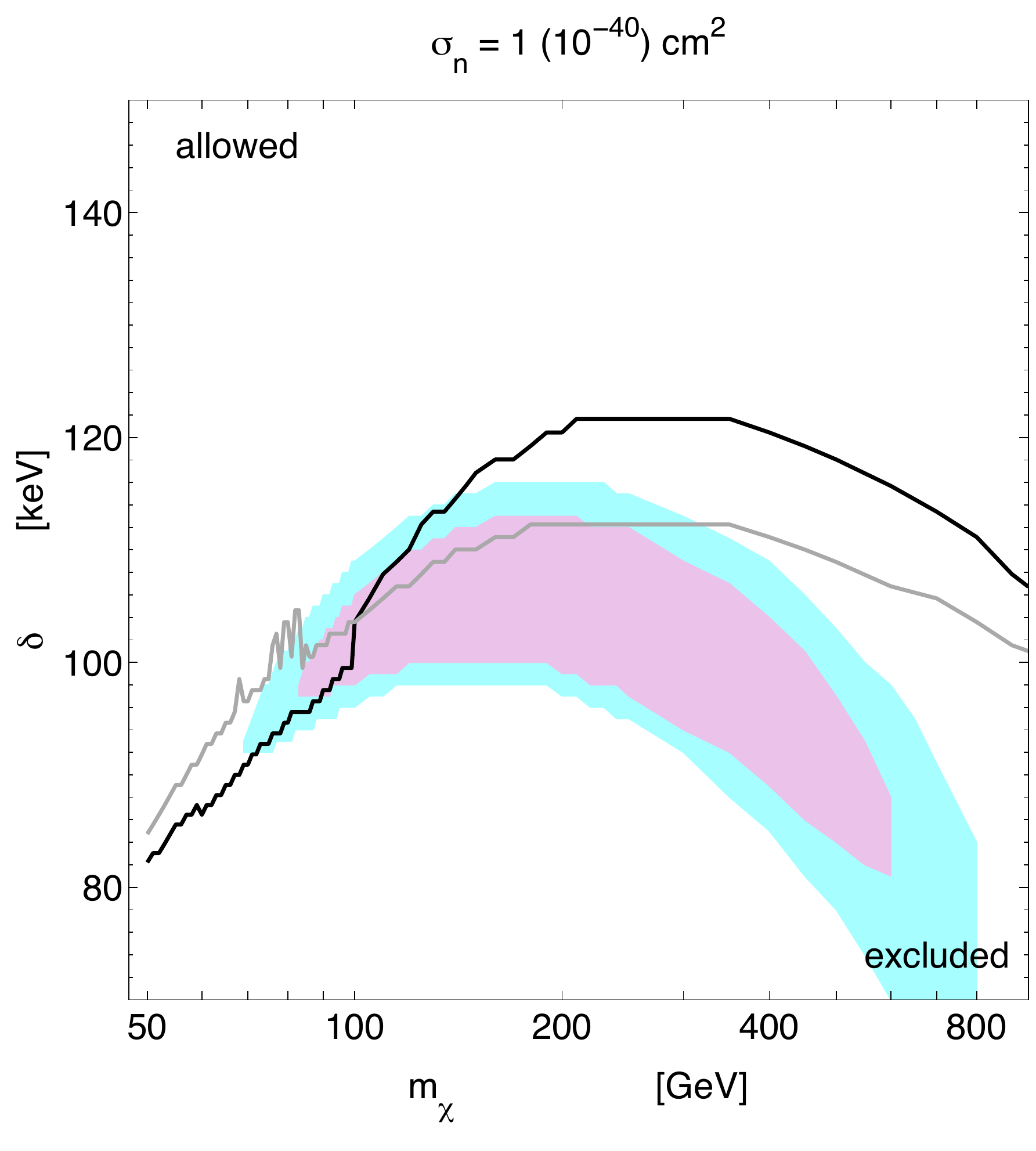}
\includegraphics[width=0.42\textwidth]{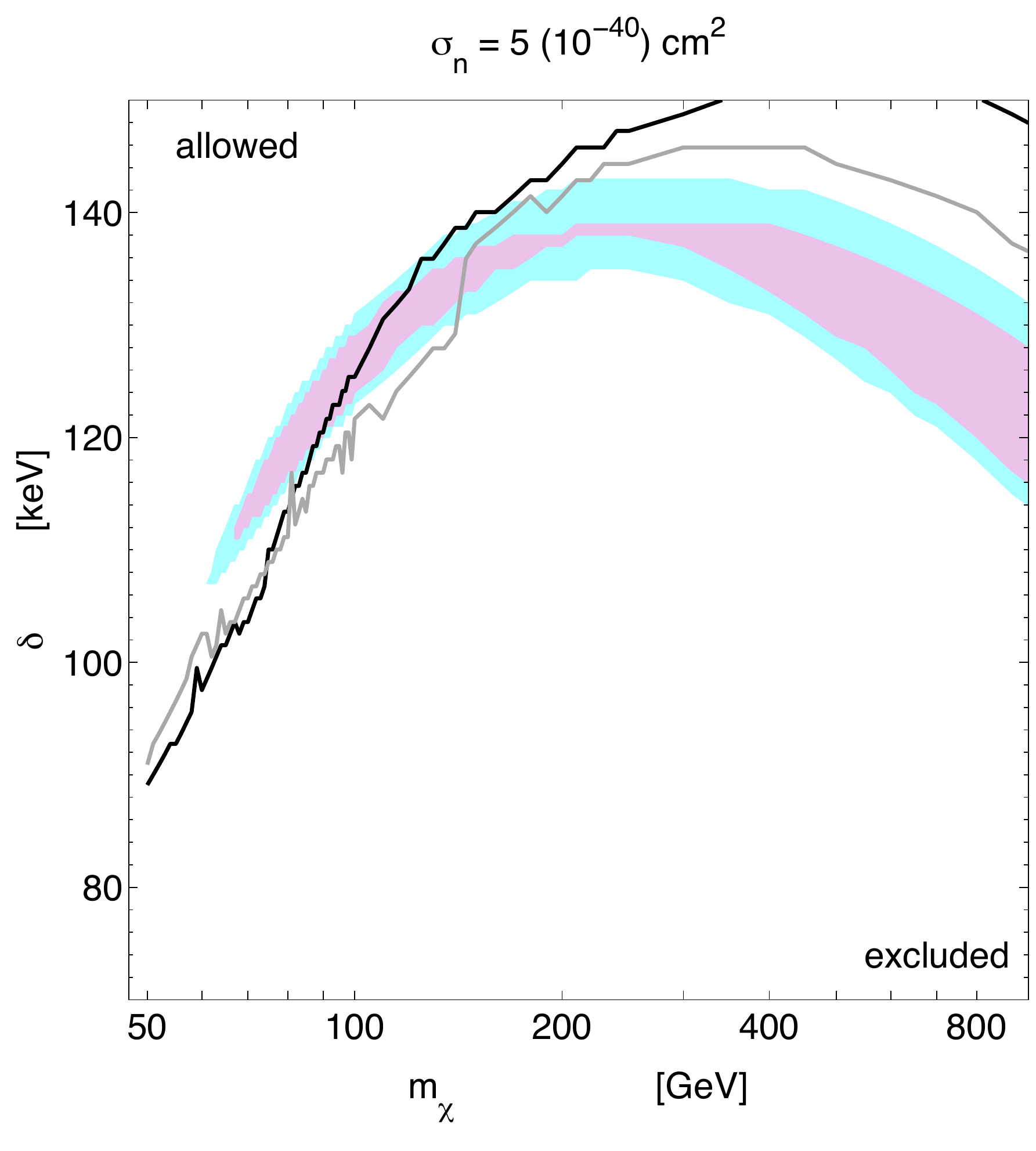}
\caption{(color online) The solid black curves indicates the 90\% C.L. exclusion limits obtained from XENON10 data, with $\mathcal{L}_{eff}$ given by \cite{2009aprile,2009sorensen} (solid gray curves take $\mathcal{L}_{eff}$ from \cite{2009manzur}).  The filled regions indicate the 90\% (magenta, darker) and 99\% (cyan, lighter) C.L. DAMA-allowed regions.}
\label{fig4}
\end{figure*}

We calculate 90\% CL exclusion limits on allowed regions of inelastic dark matter parameter space separately for $\sigma_n-m_{\chi}$, $\sigma_n-\delta$ and $\delta-m_{\chi}$, treating all 13 remaining events in the acceptance box as potential dark matter signal.  The results are shown in Fig. \ref{fig4}, for several representative cases.  We use the $p_{{Max}}$ method from \cite{2002yellin}, which explicitly provides for the presence of background events.  This allows us to place conservative exclusion limits, without background subtraction (as in \cite{2009zeplinII}, for example).  We verified that our results agree with \cite{2009chang} if we only consider events remaining in the acceptance box in \cite{2008xenon10SI}, take a constant $\epsilon \cdot A_{nr}=0.3$, and a constant $\mathcal{L}_{eff}=0.19$.  Regions of parameter space allowed by the DAMA modulation data \cite{2008dama} were calculated following the procedure described in \cite{2009chang}, and are shown as filled magenta (90\% C.L.) and cyan (99\% C.L.) regions.

Our results strongly disfavor dark matter particle masses $m_{\chi}\gtrsim150$~GeV, if we use the $\mathcal{L}_{eff}$ measurement from \cite{2009aprile,2009sorensen} (Fig. \ref{fig4}, solid black curves).  For lighter masses, our 90\% CL exclusion limits still find common parameter space with the 99\% CL DAMA-allowed region.  The behavior of the exclusion limits are similar if we use the recent $\mathcal{L}_{eff}$ measurement from \cite{2009manzur} (Fig. \ref{fig4}, solid gray curves),  however, larger dark matter particle masses remain consistent with DAMA for 110~keV~$\lesssim\delta\lesssim130$~keV.  In this case, we also find that values of $\delta\lesssim100$~keV are more strongly disfavored.  

The inelastic dark matter model is sensitive to the value of the galactic escape velocity, which we have fixed at $v_{esc}=500$~km~s$^{-1}$.  If instead we take $v_{esc}=600$~km~s$^{-1}$, dark matter particles with higher kinetic energy would remain bound in the halo.  The predicted event rate in XENON10 at higher recoil energies would thus increase, with the result that our data would more tightly constrain the DAMA-allowed parameter space.

In addition to the exclusion limits shown in Fig. \ref{fig4}, it is worth mentioning that our results can also be viewed as a limit on the modulated fraction of the dark matter signal.  A primary benefit of this quick analysis is that it requires no assumptions about the dark matter halo velocity distribution.  The sole assumption is that the DAMA modulation is due to nuclear recoils from iodine nuclei, quenched relative to electron recoils by $q_{I,nr}\simeq0.09$. Following \cite{2009chang}, the expected number of signal events observed by XENON10 can be related to the modulation amplitude observed by DAMA in the energy range $2-6$~keVee ($22-67$~keVr) according to
\begin{equation}
N_{X} = 4 k S_m (1/f-1),
\end{equation}
where the modulated amplitude is $S_m=0.0131\pm0.0016$~counts/kg/keVee as measured by DAMA \cite{2008dama}, and $f$ is the fraction of that signal that is modulated.
The prefactor $k = \epsilon \cdot A_{nr}(5.4~\mbox{kg})(58.6~\mbox{days})=92$ accounts for the total exposure obtained by XENON10.  Accounting for $\mathcal{L}_{eff}$, 12 of the 13 remaining events fall in the specified energy range, so taking the 90\% C.L. Poisson upper bound gives $N_X=18$.  From this, we find the model-independent result that a modulation $f>0.21$ is required for the two experiments to remain consistent.  Considering a $2\sigma$ downward fluctuation in $S_m$ still requires $f>0.17$.  Such large modulation arises naturally for inelastic dark matter because of the minimum kinetic energy $\delta$ that a dark matter particle must have in order to scatter.

For completeness we note that the analysis presented here does not result in a significant change to the WIMP-nucleon exclusion limit quoted in \cite{2008xenon10SI}, for the case of spin-independent elastic dark matter particle scattering.  The ``maximum gap'' method of \cite{2002yellin} was used in \cite{2008xenon10SI}.  Although the number of mis-identified events has decreased dramatically in the present work, the maximum energy interval containing no potential signal events remains similar at approximately $4.5-18$~keVr, taking (the more conservative) $\mathcal{L}_{eff}$ measurement from \cite{2009manzur}.

\section{Summary}
A primary conclusion of this work is that the events which have previously been referred to as ``anomalous'' background have been confirmed to originate in the passive layers of LXe surrounding the target, as explained in Sec. \ref{backgrounds}. We have shown that these background events can be targeted with high efficiency by the cuts described in Sec. \ref{ppc}.  A corollary conclusion is that the occurrence of this class of events can be explicitly prevented by the design of the detector, rather than by software cuts.  Such a situation is realized if there is no xenon outside the active target region, or (more realistically) if the photomultipliers viewing the target volume are perfectly  blind to scintillation produced in passive areas of liquid xenon, including below the cathode grid.  We have shown that the XENON10 dark matter search data exclude previously allowed regions of the DAMA-allowed parameter space, in the model of inelastic dark matter.  Specifically, dark matter particle masses $m_{\chi}\gtrsim150$~GeV are disfavored.  While there are events remaining in the dark matter acceptance box in Fig. \ref{fig1}, and while these events appear consistent with the expected spectral shape for inelastic dark matter, we do not claim a detection.  It was demonstrated in Sec. \ref{backgrounds} that the number of remaining events is reasonably consistent with the expected number of leakage events, based on electron recoil calibration data.  
 
\section*{Acknowledgements}
The authors would like to thank N. Weiner for insight and discussions on inelastic dark matter.  This work also benefitted from discussions during the ``New Paradigms for Dark Matter'' workshop at the University of California, Davis, December 5-6, 2008.  We gratefully acknowledge support from NSF Grants No. PHY-03-02646 and No. PHY-04-00596, CAREER Grant No. PHY-0542066, DOE Grant No. DE-FG02-91ER40688, NIH Grant No. RR19895, SNF Grant No. 20-118119, FCT Grant No. POCI/FIS/60534/2004 and the Volskwagen Foundation.

\end{document}